\documentclass{aa}
\usepackage{graphicx}
\usepackage{txfonts}
\usepackage{lscape}             
\usepackage{placeins}           
\usepackage[starfontserif 
    ]{starfont}                                
\usepackage{enumitem}
\usepackage{xcolor}
\usepackage{ulem}

\newcommand{\aox}{\ensuremath{\alpha_{\textsc{ox}}}}
\usepackage[dvipsnames]{xcolor}

\begin{document}

   \title{Weak emission line quasar SDSS J101353.45+492758.1\\ I. Continuum fitting}

   \author{L. Gibaud\inst{1}, M. Nikołajuk\inst{1}, P. Życki\inst{2}, A. Różańska\inst{2}, K. Hryniewicz\inst{3}
        \and R. Wojaczyński\inst{2}
        }

   \institute{Faculty of Physics, University of Białystok, ul. Ciołkowskiego 1L, 15-245 Białystok, Poland\\
             \email{l.gibaud@uwb.edu.pl},
            \and Nicolaus Copernicus Astronomical Center, Polish Academy of Sciences, ul. Bartycka 18, Warsaw, Poland,
            \and National Centre for Nuclear Research, Astrophysics Division, ul. Pasteura 7, 02-093 Warsaw, Poland }

   \date{Received xxx xx, 2025}

 
  \abstract
 {Weak emission-line quasars (WLQs) are active galactic nuclei (AGN) characterized by unusually faint or absent broad emission lines. A subset also exhibits pronounced X-ray weakness, offering keys insights into accretion flow structure and the physical state of the broad-line region (BLR).}
 {We present a broadband study of the WLQ SDSS J101353.45+492758.1, which displays a nearly featureless UV–optical spectrum with only a weak Mg\,\textsc{ii} line alongside an exceptionally low X-ray flux.}
 {We model its spectral energy distribution (SED) using the relativistic thin-disk model \texttt{kerrbb} with a power law, and the multicomponent AGN model \texttt{relagn}, a physically motivated extension of \texttt{agnsed} incorporating warm and hot Comptonizing regions. Our fits constrain the black hole (BH) mass, accretion rate, X-ray loudness, and coronal energetics.}
 {Both approaches yield consistent BH masses of $M_{\rm BH} \approx 2\times10^{9}\,M_\odot$ and an Eddington accretion rate of $\dot m \approx 0.1$. 
 The \texttt{relagn} fit including a warm Comptonizing region provides a significantly improved representation of the UV–soft X-ray continuum. 
 The warm corona, characterized by $kT_{e} \simeq 0.20$\,keV, $\Gamma \simeq 3.8$, and an optical depth $\tau \simeq 7.26$, extends to $\sim 34\,R_{\rm g}$. 
 The hot corona appears compact and energetically suppressed, leading to an intrinsically weak X-ray output with $\log(L_{\rm X}/L_{\rm bol}) \simeq -4.29$, among the lowest reported for WLQs. The $\aox \sim 2.06$ indicates the source to be in high/soft AGN spectral state.}
 {The combination of a luminous, standard disk and extremely weak hot corona suggests that this quasar hosts a highly inefficient inner coronal region. This explains its X-ray faintness and extreme deficit of high-ionization emission lines. The source may represent an AGN analog in "ultrasoft" accretion state, or a system in which the ionizing continuum is suppressed by a compact or quenched corona. Our study suggests that the source is not accreting at high Eddington ratio, highlighting the physical diversity of WLQs, and supports the view that geometric and radiative effects jointly shape their extreme spectral properties.}

   \titlerunning{SDSS J101353.45+492758.1}
   \maketitle

\section{Introduction}

Strong and broad emission lines are hallmarks of the optical-UV spectra of active galaxies, especially quasars. However, various sky surveys have identified objects classified as quasars that exhibit weak or even undetectable emission lines - these are known as weak emission line quasars (hereafter referred to as WLQs -- e.g., \citealt{mcdowell1995, diamondstanic2009, plotkin2010apj}). Although WLQs share many characteristics with typical quasars \citep{shemmer2009}, their optical-UV spectrum properties remain enigmatic. Their emission lines have significantly weaker equivalent widths (EW) compared to classical quasars. For instance, the EW of the CIV emission line in WLQs is smaller than or equal to 10 \r{A}, while in typical quasars its 1 $\sigma$ range spans from 26 to 67 \r{A} \citep{diamondstanic2009}. Broad emission lines can be divided into two categories \citep{wills1985, collinsouffrin1988}. The first category consists of high-ionizationlines (HILs), emitted by a highly ionized low-density region of the broad emission-line region (BLR), including lines such as C IV and He II. The second category comprises low-ionization lines (LILs), which originate from a denser, less ionized region of the BLR or from a region farther from the ionizing source. It produces, for example, the Mg II, Fe II and hydrogen Balmer lines.

WLQs, though uncommon, pose a significant challenge to our current understanding of quasar structure and physics -- particularly the nature of the BLR. Several hypotheses have been proposed to explain the physical origin of WLQs.

It was speculated that WLQs have intrinsically normal UV lines that appear weak because of dilution by a relativistically boosted continuum, similar to BL Lac objects. However, this is unlikely for most WLQs because they are radio-quiet sources \citep{plotkin2010apj}.

Some authors have proposed that WLQs trace a youthful quasar stage, where the BLR has not yet fully formed (e.g., \citealt{hryniewicz2010, andika2020}). While the accretion disk is already active and producing a standard continuum, the BLR is in progress. An alternative explanation suggested by \citet{shemmer2010} is that the weak emission lines stem from a sparsely populated BLR with a noticeable deficit of line-emitting material. Moreover, \citet{nikolajuk2012} suggested that the low covering factor of the BLR could drive such a weakness of the emission line.

The lack of sufficient high-energy ionizing photons is another direction to explore to understand the observed emission line weakness \citep{leighly2007a}. Such a soft spectral energy distribution (SED) could result from an atypical accretion rate. However, observations by \citet{shemmer2010} have shown that two high-redshift WLQs have accretion rates typical of normal quasars with comparable luminosities and redshifts (mean Eddington ratio is of 0.35, \citealt{shen2011}, while of 0.4 in Shemmer's sample). 

Lastly, one might speculate the presence of a material absorbing the ionizing photons (in the UV and X-ray energy bands) before they reach the BLR. \citet{wu2011} studied a group of X-ray weak quasars with UV emission line features similar to PHL 1811, including weak and blueshifted high-ionization lines. These PHL 1811 analogs showed X-ray luminosities on average 13 times lower than expected, along with harder X-ray spectra, suggesting strong absorption. Interestingly, these analogs may constitute nearly 30$\%$ of WLQs. While early studies (e.g., \citealt{shemmer2009, wu2011}) emphasized X-ray weakness as a defining trait of WLQs, more recent works have suggested the presence of a shielding gas that selectively blocks high-energy photons from reaching the BLR (\citealt{wu2011, luo2015, ni2018}). Such a shielding material geometry and orientation effects may play a key role. Depending on the observer's line of sight, the same object might appear as a PHL 1811 analog (if viewed through the shielding gas and BLR) or as an X-ray normal quasar. In this case, the mean ratio of the X-ray luminosity in the 0.5-10 keV range to the bolometric luminosity is of 0.034, calculated from 783 quasars, based on the catalogs from \cite{shen2011} and \citep{young2009}.

In this paper, we performed the broad band spectral analysis of the intriguing WLQ SDSS J101353.45+492758.1 (hereafter SDSS J101353). Sec.~\ref{sec:source} provides a description of the quasar and outlines the data collection. In Sec.~\ref{sec:ana}, we present the Mg II line analysis, including the measurement of the Mg II emission line. Sec.~\ref{sec:disk} is dedicated to the disk fitting method. In this section, we first detail the data corrections applied, followed by the procedure used to fit the SED of SDSS J101353. In Sec.~\ref{sec:obs}, we explore another scenario: the obscuration and the reflection of X-rays as an alternative explanation of the X-ray faintness. The results, along with a discussion, are presented in Sec.~\ref{sec:discus}, and conclusions are given in Sec.~\ref{sec:concl}.

\section{Quasar SDSS J101353.45+492758.1}
\label{sec:source}

\subsection{The source description}

With the aim of collecting a large sample of WLQs with redshift ranging from 0.19 to 3.50, we carefully examined the spectra of a dozen of WLQs from the Sloan Digital Sky Survey (SDSS) optical catalog data release 17 (DR17) \citep{abdurrouf2022}. Among these selected quasars, the spectrum of SDSS J101353 retained our attention, showing a normal quasar spectrum (see Fig.~\ref{fig1}), with almost no emission lines. Although weak, the Mg\,II emission line is clearly detected. However, the C\,IV and C\,III] emission lines are not clearly visible (see spectrum in Fig.~\ref{fig2}). 
Additionally, the quasar exhibits low level of X-ray radiation.
Typical X-ray fluxes of luminous quasars lie around $10^{-14}$ - $10^{-13}$ erg\,cm$^{-2}$\,s$^{-1}$ in the 2-10 keV energy band \citep[e.g.,][]{young2009}. By contrast, SDSS J101353 is detected at $F_{\rm X}$ \(\sim 10^{-17} \) erg\,cm$^{-2}$\,s$^{-1}$ -- i.e., $10^3$ - $10^4$ times fainter than a typical quasar. Its X-ray luminosity in the 0.5-10 keV range estimated by \cite{young2009} is 2.25 $\times 10^{41}$ erg s$^{-1}$.
SDSS J101353 is located at redshift z = 1.63558 +/- 0.00128. Its coordinates are RA = 153.47276, Dec = 49.46615 (J2000.0). We cross-checked the redshift value based on the Mg\,II line using the SDSS interactive spectrum analyzer\footnote{http://cas.sdss.org/dr17/}. Using the online cosmology calculator\footnote{https://www.astro.ucla.edu/~wright/CosmoCalc.html} developed by \citet{wright2006}, we determined the comoving distance of SDSS J101353 to 4691.20 Mpc.

\subsection{Data collection}

   \begin{figure*}[h!]
   \centering
   \includegraphics[width=0.8\linewidth]{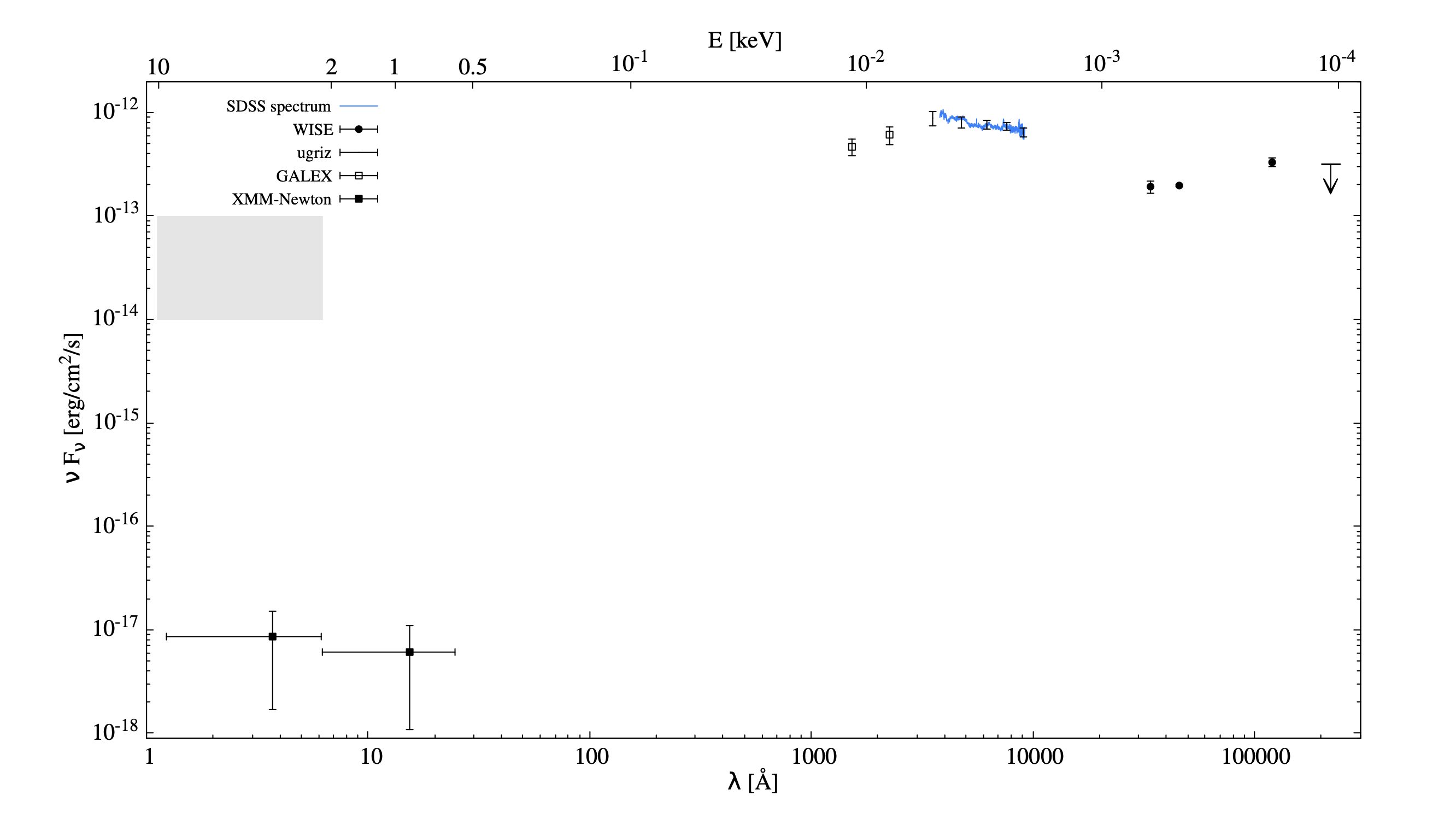}
      \caption{Data points (black) and spectrum (blue -- extracted from the SDSS data release 17) of SDSS J101353 in observed-frame. The light gray box represents the typical X-ray flux measured in the 2--10 keV energy band of classical quasars (e.g., \citealt{young2009}).}
         \label{fig1}
   \end{figure*}

To obtain the SED of SDSS~J101353, we combined data archives from data bases and catalogs. The near-IR photometric points were taken from the Wide-field Infrared Survey Explorer (WISE) satellite \citep{wright2010}. The isophotal wavelengths are 3.3526, 4.6028, and 11.5608 $\mu$m for the W1, W2 and W3 bands, respectively. The WISE W4 band (22 $\mu$m) measurement for our source is only a nonconstraining upper limit. Combined with the known challenges in calibrating W4 data, we restrict our analysis to the higher quality detections from the other bands. The optical photometric points were taken from the SDSS DR17 catalog using the five optical bands \textit{ugriz} imaging data with effective wavelengths of 3543, 4770, 6231, 7625 and 9134 \r{A}, respectively. UV data were collected from the Galaxy Evolution Explorer (Galex) satellite catalogs \citep{bianchi2011}. They consist of near-UV (NUV, 1771-2831 \r{A}) and far-UV (FUV, 1344-1786 \r{A}) data. Only two X-ray data were found from the fifth SDSS/XMM-Newton quasar survey data release \citep{young2009}. We searched for additional X-ray data using Chandra Source Catalog Release 2 \citep{evans2010} or the ROSAT all-sky survey catalog \citep{boller2016}. However, no sources with significant flux were found. 

We cross-checked our data points using the spectrum observed by SDSS and the NASA Extragalactic Database (NED)\footnote{https://ned.ipac.caltech.edu}. Fig.~\ref{fig1} shows the SED of the quasar. The spectrum has been retrieved from the SDSS data release 17.

\section{Spectrum analysis}
\label{sec:ana}
\begin{table}
\caption{Continuum and iron windows}
\label{table:1}
\centering
\begin{tabular}{c c}
\hline\hline
\noalign{\smallskip}
 & Fitting windows  \\
 & (rest-frame wavelengths in \r{A}) \\
 \noalign{\smallskip}
\hline
\noalign{\smallskip}
Continuum & 1455-1470, 1690-1700, 2160-2180  \\
          & 2225-2250, 3010-3040, 3240-3270 \\
 \noalign{\smallskip}
\hline
\noalign{\smallskip}
Iron & 2020-2115, 2250-2650, 2900-3000 \\
\noalign{\smallskip}
\hline
\end{tabular}
\end{table}

\subsection{Iron decontamination}

Active galactic nuclei (AGN), including quasars, exhibit rich iron emission due to the presence of both atomic and ionic iron -- a stable and abundant product of stellar nucleosynthesis. Because iron atoms have a highly complex electron structure with many energy levels, they produce thousands of emission lines spanning the UV and optical ranges. These numerous, often weak, transitions overlap heavily, creating a blended emission known as a pseudo-continuum \citep{vestergaard2001}. This pseudo-continuum lies above the true, intrinsic, underlying continuum and can obscure or distort weaker noniron spectral features. As a result, it introduces significant uncertainties in the measurement of both continuum placement and line fluxes.
To ensure reliable spectral measurements, the iron emission must be modeled and subtracted. Fitting and removing the iron pseudo-continuum is therefore a critical step.

We corrected the spectrum for Galactic reddening using a standard extinction law \citep{fitzpatrick1999}, parameterized by the V-band extinction $A_V$ and the total-to-selective extinction ratio \( R_V \equiv A_V / E(B-V) \). In the diffuse interstellar medium, $R_V$ typically ranges from 2.6 to 5.5, with a mean value of 3.1 \citep{cardelli1989,gaskell2004}.
The color excess $E(B-V)$ for SDSS J101353 was determined from the dust reddening map \citep{schlegel1998,schlafly2011}, and is of $0.0081 \pm 0.0009$ mag. Then,
we modeled the spectrum of SDSS J101353 by fitting a power law to the continuum and accounting for blended iron emission. The underlying continuum was fitted to spectral regions free from emission lines and unaffected by blended iron emission, called the continuum windows. These continuum windows were taken from \citet{kuraszkiewicz2002} and are listed in Tab.~\ref{table:1}. The continuum was then subtracted from the spectrum.
Next, following the method outlined in \citet{vestergaard2001}, we used their UV iron template to model the blended iron emission, applied to the spectral windows were iron emission is strong \citep{kuraszkiewicz2002}. The iron windows are listed in Table \ref{table:1}. The iron template was broadened by convolution with Gaussian functions, with widths ranging from 900 to 6000\,km\,s$^{-1}$ in increments of 250\,km\,s$^{-1}$ (while ensuring that the total flux of each template was conserved). The broadening value that minimized the $\chi^{2}$ statistic was selected to optimally fit the spectral region around the Mg II emission line, where the contribution of iron is stronger. This approach ensures a better Mg II emission line fitting, and therefore a more accurate estimation of the quasar's fundamental parameters.
We found a value of the iron full width at half maximum (FWHM) of 2900 $\pm$ 250 $\mathrm{km}\ \mathrm{s}^{-1}$. The best-fit iron emission was subtracted from the quasar's spectrum.

\begin{figure*}[t]
\centering
  \includegraphics[width=1.0\linewidth]{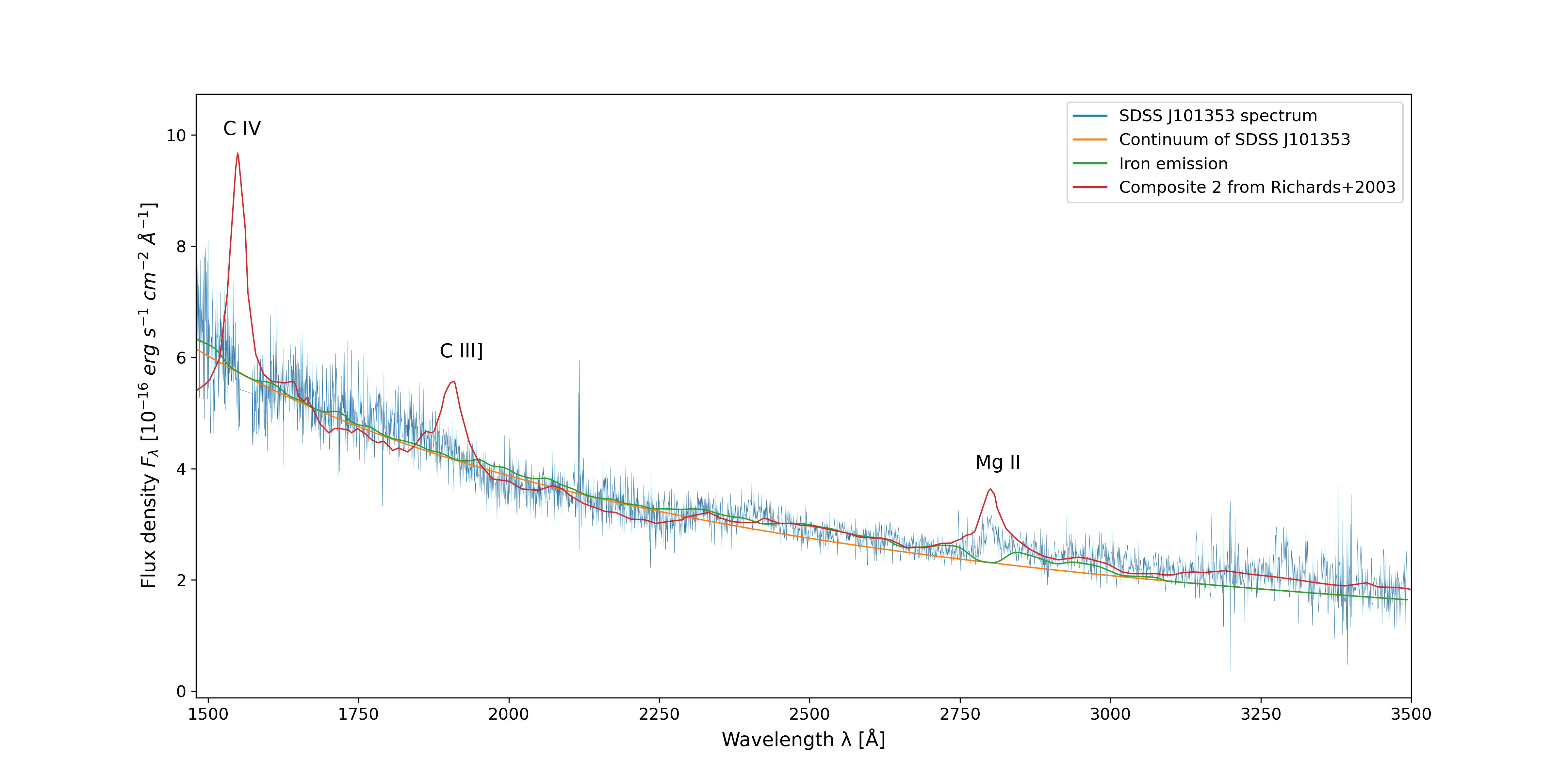}
    \caption{Rest-frame spectrum of SDSS J101353 (blue) shown with the fitted continuum (orange) and the broadened iron emission (green). For comparison, the composite quasar spectrum 2 from \protect\citet{richards2003} is also displayed (red).}
    \label{fig2}
\end{figure*}

We found that the best-fit continuum has a power-law index \( \alpha_{\lambda} = -1.54 \pm 0.01 \), defined such that the flux density \( F_\lambda \propto \nu^{\alpha_\lambda} \). This corresponds to \( \alpha_{\nu} = -0.46 \), where \( F_\nu \propto \nu^{\alpha_\nu} \). Comparing the continuum of SDSS J101353 with the quasar composite spectra created by \citet{richards2003}, we found that its continuum is between the composites no. 2 and no. 3 for which the spectral indexes \(\alpha_{\nu}\) (measured from 1450 to 4040 \r{A}) are -0.41 and -0.54, respectively. The rest-frame spectrum of SDSS J101353, extracted from the SDSS data release 17, is plotted in Fig.~\ref{fig2}, along with the continuum and the iron emission fits. The quasar composite spectrum no. 2 from \cite{richards2003} is also displayed. While their continua are broadly similar, the two spectra differ significantly in their emission line features. The Mg II emission line is clearly visible, though notably weaker in the SDSS J101353 spectrum. In contrast, the semi-forbidden Si III]+C III] line and the C IV emission are not easily visible from the WLQ spectrum.

\subsection{Mg II emission line fitting and black hole mass estimation}

\begin{figure}[h!]
   \centering
   \includegraphics[width=\hsize]{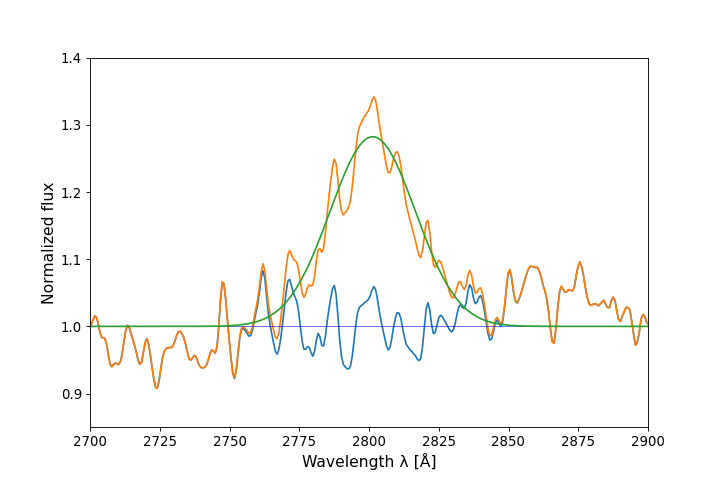}
      \caption{Mg II emission line fitting. The rebinned rest-frame spectrum of SDSS J101353 after iron subtraction is displayed in orange. The signal from which the continuum has been subtracted, underlying the emission line is shown in blue. The green line materializes the best fit with a single Gaussian.}
         \label{fig3}
   \end{figure}

Black holes (BHs) masses in AGN can be estimated through several complementary techniques. The most direct approach is the reverberation mapping, which measures time delays between continuum and emission-line variations to infer the BLR size and, hence, the virial mass. When such monitoring data are unavailable, single-epoch methods based on broad-line widths and empirical scaling relations provide a reliable alternative. Therefore, we used the Mg II emission line (commonly referred to as Mg II $\lambda 2800~\AA$) to determine the central BH mass of SDSS~J101353. 
The Mg II feature is a resonance doublet at $\lambda 2797~\AA$ and $\lambda 2803~\AA$. In broad-line AGN, the typical line width (a few thousand km $\text{s}^{-1}$) exceeds the separation of the two components (corresponding to a velocity difference of $\sim 770 ~\text{km s}^{-1}$), so they are blended and treated as a single feature near $2800~\AA$. 
This approximation is widely adopted in single-epoch virial mass estimators and has a negligible impact on the inferred BH mass. 
The line was fitted using a Gaussian profile, as shown in Fig.~\ref{fig3}. The results of the fit are summarized in Tab.~\ref{table:2}. We found a rest frame EW of \( 10.89 \pm 0.57~\AA \). 
To verify the robustness of the derived line parameters, we tested whether the fit depends on the data binning. We found that the FWHM remains consistent within the fitting uncertainties for different binning methods, and therefore its value can be reliably used for the BH mass estimation. 
The monochromatic luminosity at 3000 $\AA$ taken from the SDSS spectrum of SDSS J101353 is \( \lambda L_{\lambda} (3000~ \AA) = 2.07 \times 10^{45} \) erg $\mathrm{s}^{-1}$. This luminosity is typical for a quasar \citep{dempsey2018}. We used the bolometric correction at 3000 $ \AA$ from \cite{shen2011}, \( BC_{3000} = 5.15 \) and obtained a bolometric luminosity of \( 1.06 \times 10^{46} \) erg $\mathrm{s}^{-1}$. We calculated the mass of the BH (in units of solar mass $\text{M}_\sun$) using equation (1) from \cite{vestergaard2009}:
\[
 \text{M}_{\text{BH}} = 10^{6.86}
 \left[ \frac{\text{FWHM}(\text{Mg II})}{1000~ \text{km/s}} \right]^2
 \left[ \frac{\lambda L_{\lambda}(3000~ \AA)}{10^{44}~ \text{erg/s}} \right]^{0.5} 
\]
We found a mass of \( (4.95 \pm 0.64) \times 10^8\ \text{M}_\sun \), consistent with those found in the literature, in particular with those from \cite{vestergaard2008} (their figure 1).\\
Knowing the BH mass, the Eddington accretion rate can be determined with the relation: \(\dot m = L_{\rm bol}/L_{\rm Edd}\), where $L_{\rm bol}$ is the bolometric luminosity and $L_{\rm Edd}$ is the Eddington luminosity \citep{rykickilightmann1979}.
The accretion rate of our WLQ is 0.17.

Among the sample of about 300 type-1 quasars studied by \cite{sulentic2006}, quasars with similar redshifts have a mass ranging from \( \sim 6 \times 10^8 \) to \( \sim 10^{10}\ ~\text{M}_\sun \) (see their figure 5) and show an accretion rate ranging from $\sim$0.15 to $\sim$1.20 (see their figure 7).

\begin{table}
\caption{Rest-frame Mg II properties of SDSS J101353}
\label{table:2}
\centering
\begin{tabular}{c c c}
\hline\hline
\noalign{\smallskip}
$\lambda$ &          FWHM          &   rest EW   \\
 (\r{A})  & (km $\mathrm{s}^{-1}$) & (\r{A}) \\
\noalign{\smallskip}
\hline
\noalign{\smallskip}
$2801.63 \pm 1.31$ & $3875 \pm 250$ & $10.89 \pm 0.57$ \\
\noalign{\smallskip}
\hline
\end{tabular}
\tablefoot{Statistical errors only.}
\end{table}

\section{Disk fitting method}
\label{sec:disk}

Another method of estimating the mass and the accretion rate of an accreting source is to model its broadband SED.
Therefore, we fit the broadband SED of SDSS J101353 using the X-ray Spectral fitting package XSPEC \citep[version 12.13.1,][]{arnaud1996}. 
The optical-UV bump is attributed to the emission generated by the accretion disk surrounding the central engine. The X-ray data points are interpreted as emission from a Comptonizing reservoir such as, e.g., a hot corona and a warm region. A file containing additional optical-UV continuum photometric data, besides \textit{ugriz}, was created from the SDSS spectrum file by selecting the continuum windows of \cite{kuraszkiewicz2002} listed in Table \ref{table:1}. The SDSS spectrum was binned and converted into XSPEC compatible files. For this step, we used the tool \texttt{ftflx2xsp}. The generated SED file used in XSPEC contains 20 data points.

Corrections are necessary for the data because they are tainted by various factors. We explain these corrections first in the next subsection. The spectral analyses using XSPEC are detailed further below.

\subsection{Data corrections}

We corrected our observed data for Galactic reddening using the command \texttt{redden}. We used the color excess E(B-V) of $0.0081 \pm 0.0009$ mag, that was previously determined for SDSS~J101353 from the dust reddening map \citep{schlegel1998,schlafly2011}.

We also applied a photoelectric absorption correction caused by the Galactic and intergalactic medium using the command \texttt{phabs}. This photoelectric absorption influences in particular the X-ray fluxes. We used the measurement of the Galactic neutral atomic hydrogen column density, $N_H$, derived from the HI 4$\pi$ all-sky survey \citep{hi4pi2016}, which is of $9.15\times 10^{19}\ \mathrm{cm}^{-2}$. In all the fitted models, the color excess, $N_H$, the redshift and the distance of the quasar are fixed.

\subsection{Starlight and torus contributions}

Another factor to consider is the contribution of starlight to the SED from the stars of the galaxy hosting the quasar. It is often suggested that this contribution is negligible \citep{shen2011}. However, we decided to include it. We tested different templates of \cite{polletta2007}. Assuming that the galaxies hosting quasars are elliptical, we used in particular the template of a 5 giga-year-old elliptical galaxy that matches our redshift. We created this component in the form of a readable XSPEC file due to the absence of such a model in this package. Adding a starlight component to our SED gives better fit results, especially for the infrared data.

As the WISE data show evidence of a dusty torus surrounding the central engine and its accretion disk, we added one black body component to our SED (\texttt{bbody} model in XSPEC) as thermal emission from the torus. The mean temperature obtained is $930 \pm 100$ K. The typical hot thermal dust emission in WLQs ranges from 870 K to 1240 K according to \cite{diamondstanic2009}. This is close to the typical emission of host dust in type-1 quasars, ranging from 1100 K to 2200 K, according to \cite{collinson2017}.

\subsection{Fitting using a multi-temperature blackbody model and a power law}

\begin{figure}[h!]
   \centering
   \includegraphics[width=\hsize]{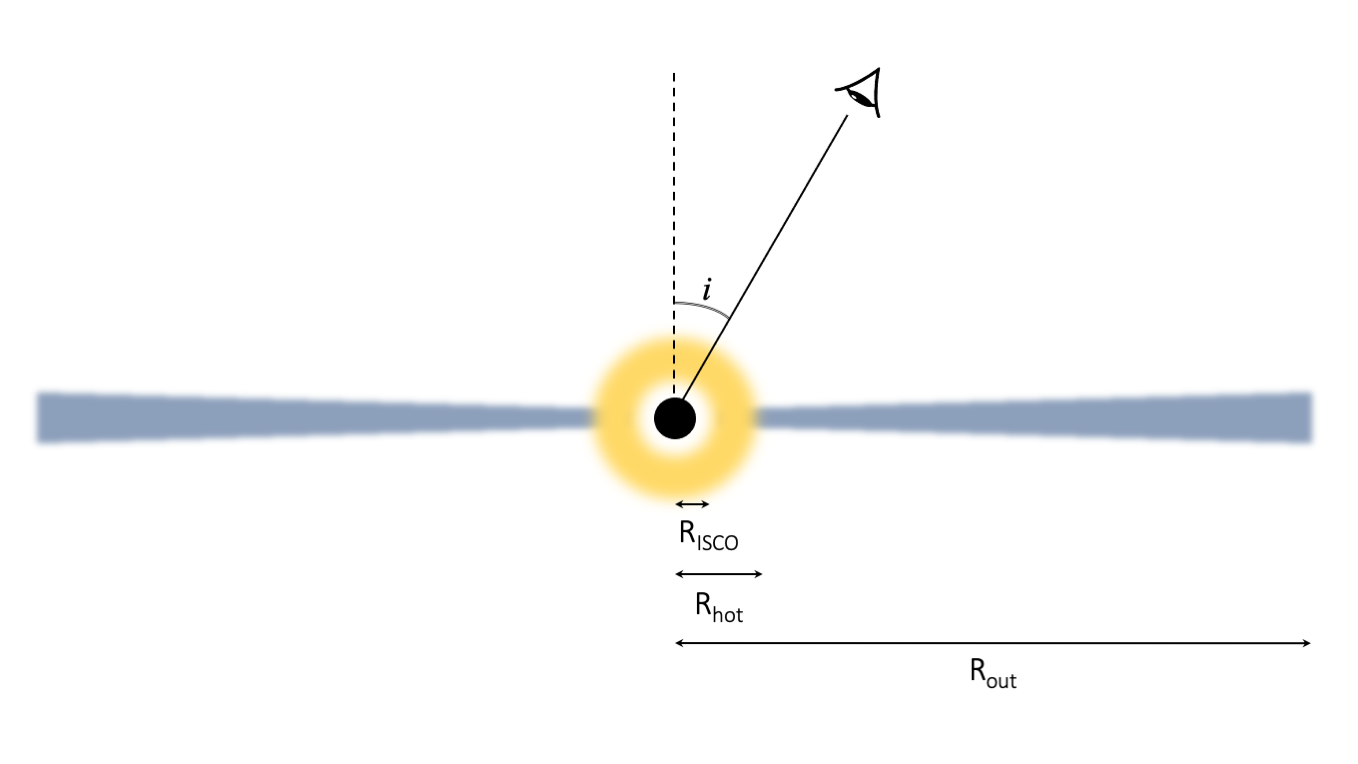}
      \caption{Schematic geometry used in the fitting process for Models 1a. The BH is in black, the standard geometrically thin accretion disk in blue-gray, and the hot corona in yellow.}
         \label{fig4}
   \end{figure}

We fitted the available photometric points with a minimal physical model consisting of a relativistic multi-color blackbody model using the \texttt{kerrbb} component \citep{li2005} in XSPEC to represent the accretion disk emission, along with a simple power-law component intended to model the X-ray emission (hereafter Model 1a, illustrated in Fig. \ref{fig4}).

All additive components of Model 1a are corrected for the quasar's redshift. This model is described as follows:
\begin{equation}\label{1}
\texttt{redden} \times \texttt{phabs} \times (\texttt{kerrbb} + \texttt{powerlw} + \text{starlight} + \texttt{bbody})
\end{equation}
The relativistic model \texttt{kerrbb} describes the emission of a thin disk around a BH described by a Kerr metric. The key assumption of the \texttt{kerrbb} model is that all the heat generated within the accretion disk is instantaneously radiated away, with no significant energy advection or internal storage. This model was originally developed to describe the thermal emission from accretion disks around stellar-mass BHs, particularly in X-ray binaries. However, in our study, we extend its application to supermassive BHs.

\begin{figure}[h!]
   \centering
   \includegraphics[width=\hsize]{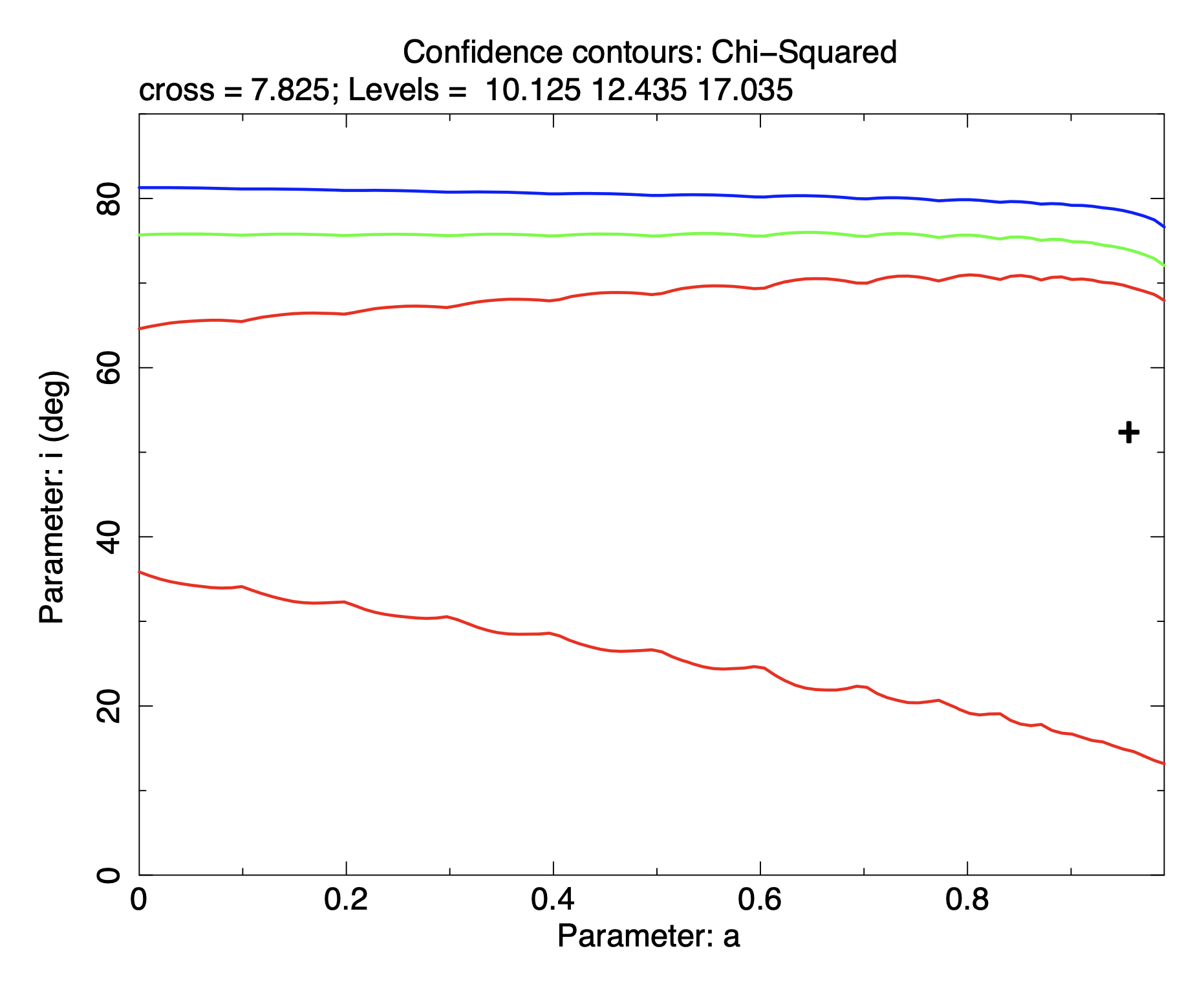}
      \caption{$\chi^{2}$ confidence contours in the spin-inclination ($a$–$i$) plane, using Model 1a: \texttt{kerrbb} + power law components. This was obtained by allowing the BH mass and mass accretion rate to vary. The red, green, and blue contours correspond to the 68\%, 90\%, and 99\% confidence levels, respectively. The cross indicates the best-fit solution.}
         \label{fig5}
   \end{figure}

\begin{figure}[h!]
   \centering
   \includegraphics[width=\hsize]{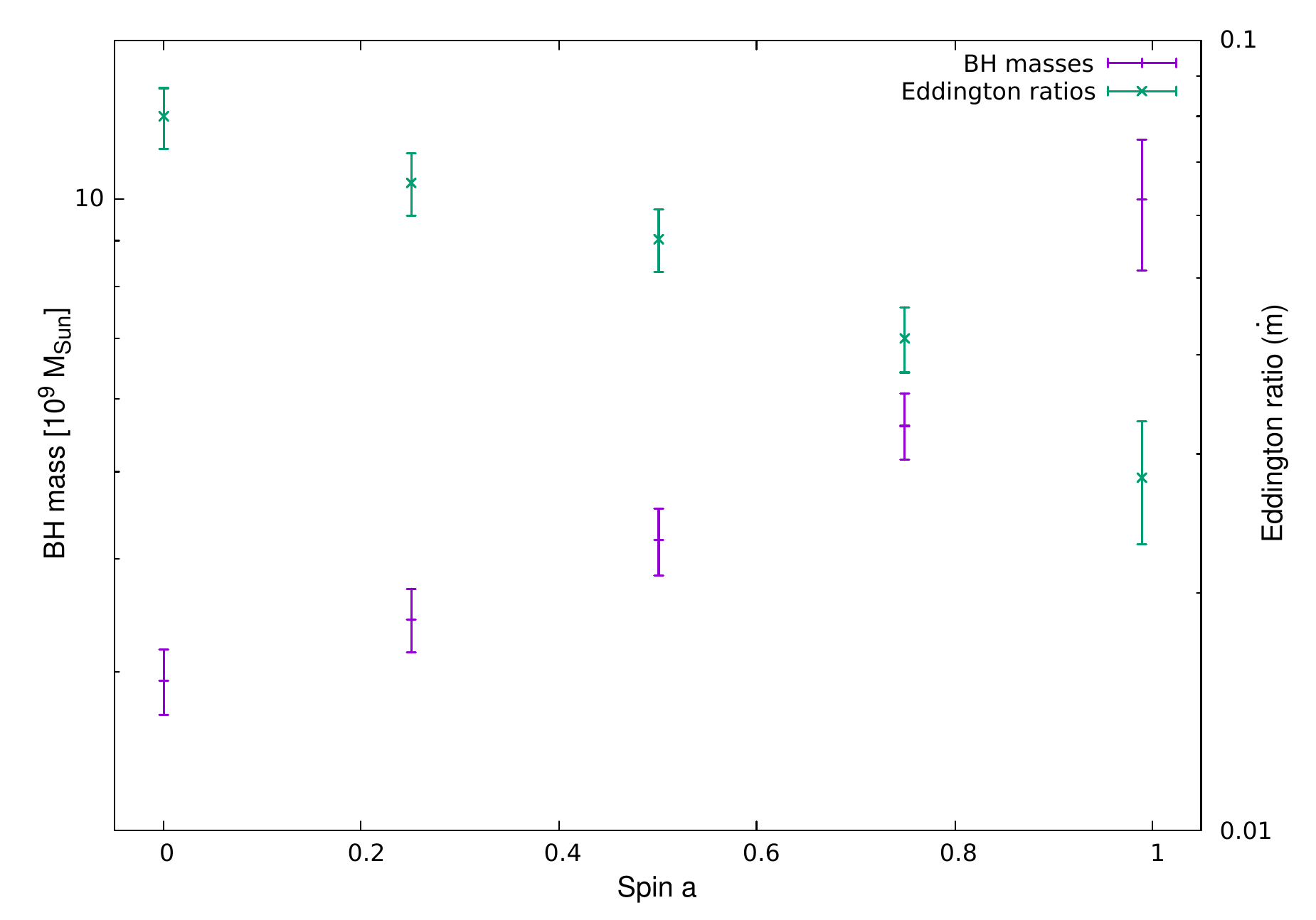}
      \caption{Dependence of the inferred BH mass and Eddington ratio as a function of the adopted spin parameter $a$, calculated for Model 1a, with an inclination $i = 50^\circ$.}
         \label{fig6}
   \end{figure}

The starlight and \texttt{bbody} components represent the starlight contribution from the host galaxy and the dusty torus, respectively (as explained in the previous subsection).
Since supermassive BHs in the centers of quasars are expected to possess non-zero spin, we tested whether our data can constrain the spin of SDSS J101353. To do this, we generated $\chi^{2}$ confidence contours in the spin-inclination plane, allowing the BH mass and the mass accretion rate to vary freely. 
The confidence contours (Fig. \ref{fig5}) indicate a mild preference for high-spin values. A nonrotating BH remains allowed at the 99\% confidence level. The inclination is best constrained around 50$^{\circ}$, with a 68\% confidence range of approximately 45$^{\circ}$-60$^{\circ}$. 
We also examined how the inferred BH mass and accretion rate depend on the assumed spin value. The results are shown in Fig. \ref{fig6}. 
Overall, the spin remains unconstrained. Adopting a zero-spin configuration therefore provides a reasonable baseline for the disk modeling.
We fixed the torque parameter (defined as the ratio of the disk power produced by a torque at the disk inner boundary to the disk power arising from accretion) to zero, corresponding to a standard Keplerian disk configuration around a nonrotating BH. 
We assumed a disk's inclination angle (that is, the angle between the axis of the disk and the line of sight) to 30$^{\circ}$, consistent with the properties of Type 1 quasars. However, we tested it, modifying the inclination angle gives results that remain in a reasonable range, as shown in Table \ref{table:3}. We decided to keep the value of 30$^{\circ}$ as a reference value for the next more sophisticated models.

\begin{table}
\caption{\texttt{kerrbb} parameters results for different values of the inclination angle \textit{i}}
\label{table:3}
\centering
\begin{tabular}{c c c c c}
\hline\hline
\noalign{\smallskip}
\textit{i} &   $\text{M}_\text{BH}$ &   Accr. rate    & Photon-index & $\chi^{2}$ / dof \\
           & ($10^9$ $\text{M}_\sun$) & ($\text{M}_\sun$/year) &    &    \\
\noalign{\smallskip}
\hline
\noalign{\smallskip}
15$^{\circ}$ & 1.76 $\pm$ 0.17 & 5.83 $\pm$ 0.24 & 1.79 $\pm$ 0.79 & 10.79 / 13 \\
30$^{\circ}$ & 2.05 $\pm$ 0.20 & 6.53 $\pm$ 0.27 & 1.79 $\pm$ 0.79 & 10.31 / 13 \\
45$^{\circ}$ & 2.63 $\pm$ 0.26 & 8.13 $\pm$ 0.33 & 1.79 $\pm$ 0.79 & 9.88 / 13 \\
\noalign{\smallskip}
\hline
\end{tabular}
\end{table}

\begin{figure}[h!]
   \centering
   \includegraphics[width=\hsize]{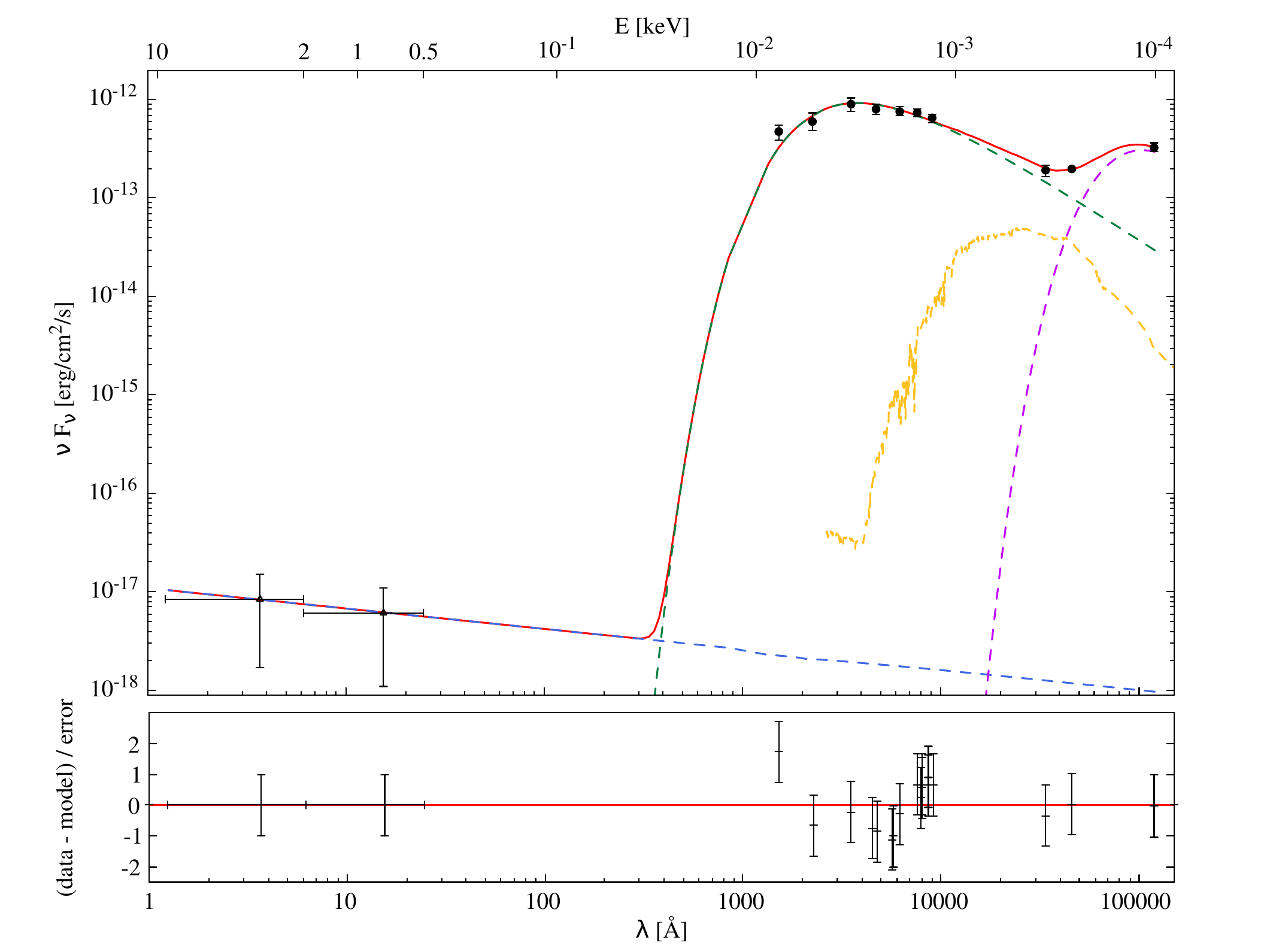}
      \caption{Broadband SED fit of SDSS J101353 using Model 1a: \texttt{kerrbb} + power law components. \textit{Top panel}: best fit model (red line) to the data (black dots). The green and blue dashed lines represent \texttt{kerrbb} and power law components, respectively. The starlight and torus contributions are shown in yellow and violet, respectively. \textit{Bottom panel}: residuals of $\chi^{2}$.}
         \label{fig7}
   \end{figure}

The best-fit parameters are summarized in Table \ref{table:4} and the broadband SED fit is shown in Fig. \ref{fig7}. The fit quality ($\chi^{2}$/dof=10.31/13) indicates that this simple model is sufficient to describe the data within the observational uncertainties. The fundamental parameters, BH mass and Eddington accretion rate, are typical to normal quasars \citep{shen2011}. 
The results show that the source emission in hard X-rays is very weak, $\approx 10^{-5}$ of the total luminosity. The fitted photon index, $\Gamma = 1.8 \pm 0.8$, is consistent with typical AGN values ($\Gamma \approx 1.7 - 2.5$). This suggests either an intrinsically weak hot corona, or the presence of a partial obscuration/shielding of the corona, so that only reflected X-rays reach the observer.

\begin{figure}[h!]
   \centering
   \includegraphics[width=\hsize]{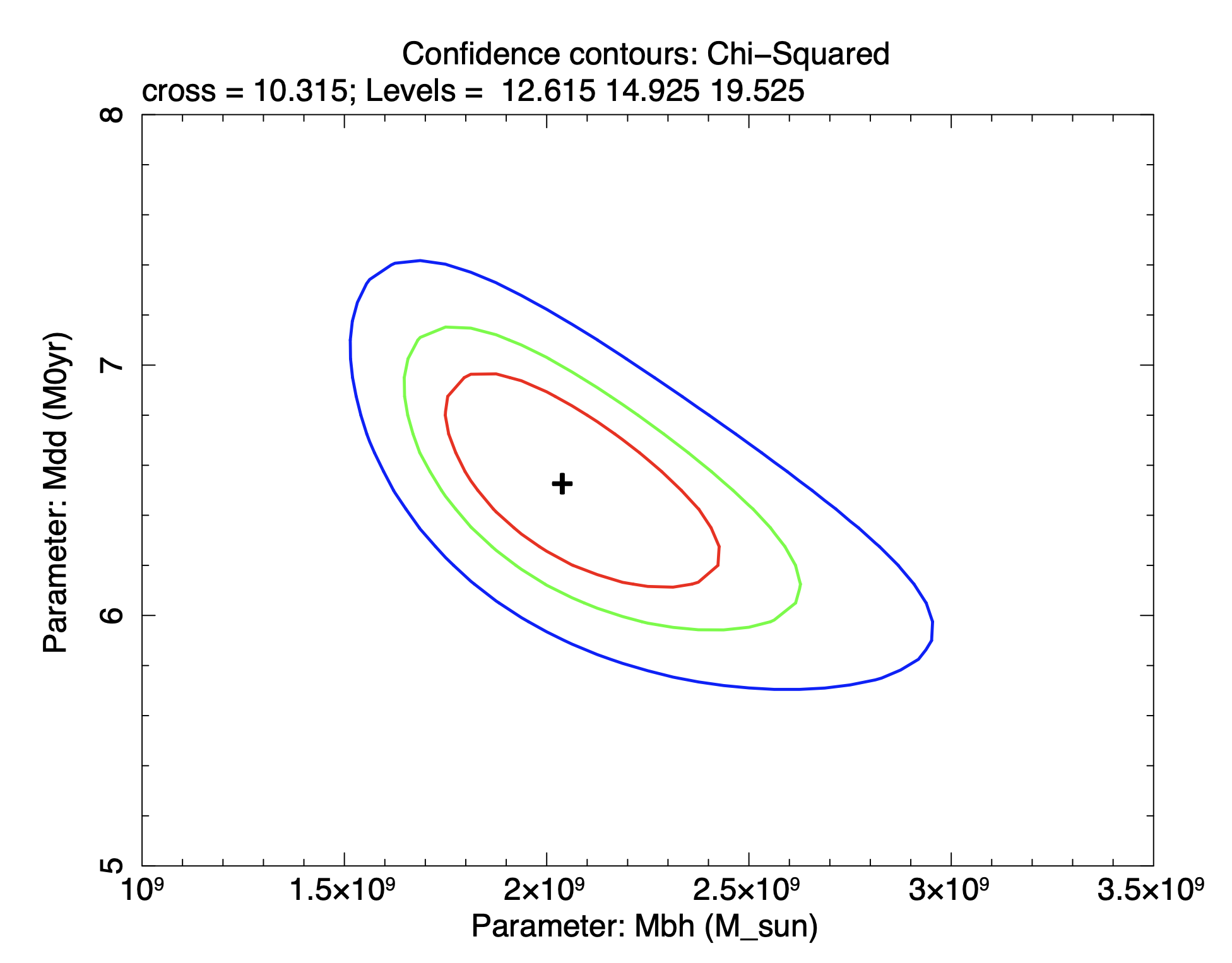}
      \caption{Confidence contours for the BH mass and the mass accretion rate for Model 1a. The red, green, and blue lines correspond to the 68\%, 90\%, and 99\% confidence levels, respectively. The cyan cross marks the best-fit values.}
         \label{fig8}
   \end{figure}

We performed a stepping investigation on the BH mass (parameter Mbh) and the mass accretion rate of the disk (parameter Mdd) to test the robustness of the fit and to explore possible parameter degeneracies. Fig. \ref{fig8} displays the $\chi^{2}$ confidence contours between the BH mass and the mass accretion rate obtained from this analysis. The red, green, and blue contours correspond to the 68\%, 90\%, and 99\% confidence levels, respectively.
The elongated shape of the contours reveals a degeneracy between the two parameters: larger BH masses require lower accretion rates to reproduce the same continuum shape. This trend reflects the intrinsic coupling between these parameters in thin-disk models, as both govern the disk temperature distribution and normalization.
The best-fit solution (cyan cross) corresponds to \( \text{Mbh} \approx 2 \times 10^{9} ~\text{M}_\sun \) and \( \text{Mdd} \approx 6.5 ~\text{M}_\sun ~\text{yr}^{-1} \). These values imply an Eddington ratio ($\dot{m} \approx 0.087$) consistent with a moderately accreting quasar, supporting the physical plausibility of the fit within the standard thin-disk framework.
The Eddington accretion rates inferred from the single-epoch method (Sec. 3.2) and from the SED fitting differ by some factor. Such differences are expected, as single-epoch virial estimates rely on broad-line widths and are therefore sensitive to uncertainties in the BLR geometry and dynamics, whereas the SED-based method provides a BLR-independent estimate of the accretion flow. In particular, the FWHM measured from single-epoch spectra may be affected in WLQs, leading to biased virial mass and accretion rate estimates \citep[e.g.,][]{mejia-restrepo2018, marculewicz2020}.

\subsection{Fitting using an AGN SED model}

\begin{figure*}
\centering
  \includegraphics[width=18cm]{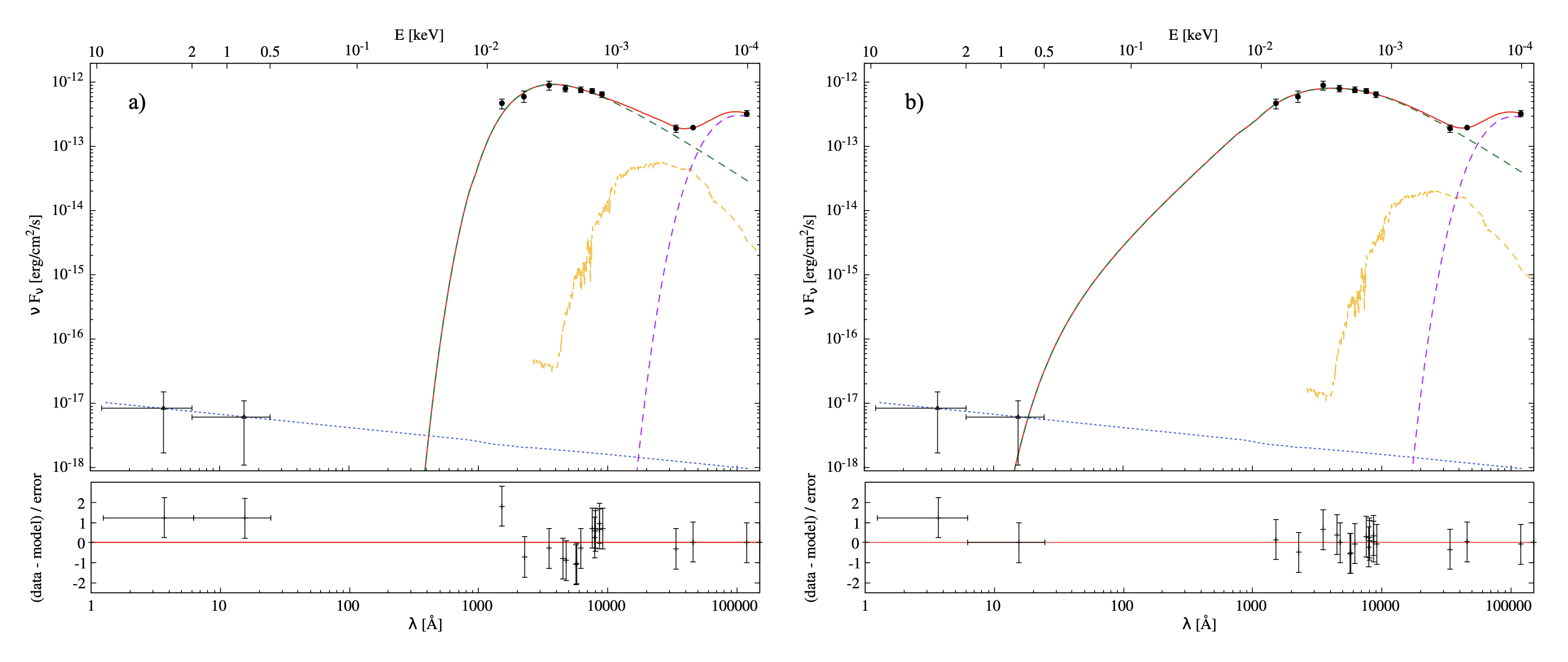}
    \caption{Broadband SED fit of SDSS J101353 using \texttt{relagn} model: plot a) represents an accretion disk with a very weak and compact hot corona configuration ; in plot b) we adopte the same configuration as in a), but letting the warm region parameters free. \textit{Top panel of each plot}: best fit model (red line) to the data (black dots). The green dashed line represents the \texttt{relagn} component. As previously, the starlight and torus contributions are shown in yellow and violet, respectively. An artificial power-law (dotted blue line) is displayed to show the contribution of a compact weak hot corona. \textit{Bottom panel of each plot}: residuals of $\chi^{2}$ related to the total model in red (without taking into account the artificial power law).}
    \label{fig9}
\end{figure*}

\begin{table*}
    \centering
    \caption{Fitting results of SDSS J101353 using different models and assuming an inclination angle of 30$^{\circ}$ from the accretion disk axis}
    \label{table:4}
    \begin{tabular}{c c c c c c c c c}
    \hline\hline
    \noalign{\smallskip}
         & Model & $\text{M}_\text{BH}$  & \(\dot{m}\)\tablefootmark{a} & $\Gamma_\text{hot}$\tablefootmark{b} & $\Gamma_\text{warm}$\tablefootmark{c} & Starlight & $kT_\text{e,torus}$ & $\chi^{2}$/dof \\
         &       & ($10^9~ \text{M}_\sun$) &  &  &  & ($10^{-17}$)  & (K)  & \\
         \noalign{\smallskip}
         \hline
         \noalign{\smallskip}
         & Model 1a & \(2.05_{-0.30}^{+0.37}\) & \(0.087_{-0.006}^{+0.005}\) & $1.8\pm0.8$ & - & $5.18\pm3.93$ & $934.16\pm98.64$ & 10.31/13 \\
         \noalign{\smallskip}
         & Model 1b & \(1.51_{-0.19}^{+0.22}\) & \(0.174_{-0.029}^{+0.035}\) & 1.8 fixed & - & $5.31\pm3.93$ & $931.84\pm97.48$ & 12.38 / 15 \\
         \noalign{\smallskip}
         & Model 2 & \(1.90_{-0.26}^{+0.28}\) & \(0.155_{-0.025}^{+0.027}\) & 1.8 fixed & $3.92\pm0.56$ & $1.83\pm0.87$ & $913.27\pm30.18$ & 3.69 / 12 \\
         \noalign{\smallskip}
         \hline
    \end{tabular}
    \tablefoot{Model 1a: \texttt{kerrbb} + power law. Model 1b: accretion disk + compact weak hot corona using \texttt{relagn}. Model 2: warm region contribution to Model 1b using \texttt{relagn}.\\
    \tablefootmark{a}{Accretion rate \(\dot{m}=\dot{M}/\dot{M_{\rm Edd}}\).}
    \tablefootmark{b}{Photon-index in the 1-10 keV energy range.}
    \tablefootmark{c}{Photon-index in the 0.1-1 keV energy range.}}
\end{table*}

While \texttt{kerrbb} provides a relativistic model of the disk emission around a BH, assuming local energy dissipation and purely thermal radiation, it does not include other important emission components known to shape AGNs continua, such as Comptonization in warm and hot corona. To model these features, we adopt the model \texttt{relagn} \citep{hagen2023}, a relativistic extension of the \texttt{agnsed} framework \citep{kubota2018}. \texttt{relagn} retains the physical three-zone structure of \texttt{agnsed} -- a standard outer disk, a warm Comptonizing region responsible for the soft excess, and a hot corona producing the X-ray power law -- while incorporating full relativistic effects on the disk emission. In both \texttt{agnsed} and \texttt{relagn}, all radiative power originates from gravitational energy released in the accretion disk. A fraction of this energy is locally diverted to power the warm and hot Comptonizing regions, while the remainder is emitted as thermal disk radiation. Thus, the warm and hot coronae do not constitute independent energy sources but reprocess disk dissipation within their respective radial zones. This makes \texttt{relagn} a more appropriate model for modeling the full broadband SED of quasars.

The strength of both the hot and warm coronae are parameterized by their outer radius, $R_{\rm hot}$ and $R_{\rm warm}$, respectively, with their minimum values at the innermost stable circular orbit (ISCO) indicating no corona at all. 
In all the fits performed, we fixed the outer radius of the accretion disk $R_{\text{out}}$ to $10^{5}~ R_\text{g}$, the upper limit of the scaleheight for the hot component to the default value, 10 $R_\text{g}$, and we did not consider reprocessing mechanisms. We assumed a nonspinning BH, so ISCO is at $6\,R_{\rm g}$.
As previously, we fixed the inclination angle (between the axis of the disk and the line of sight) to 30$^{\circ}$.
We fitted the broadband SED of SDSS J101353, focusing on the possible contribution of a hot corona and a warm region. For that, we tested two configurations with the following model:
\begin{equation}\label{2}
\texttt{redden} \times \texttt{phabs} \times (\texttt{relagn} + \text{starlight} + \texttt{bbody})
\end{equation}
%
We first attempted to fit $R_{\rm hot}$, meaning the strength of the hot corona, assuming no warm corona, $R_{\rm warm}$ fixed at $6\,R_{\rm g}$. The best fit was obtained for $R_{\rm hot} = 6 R_{\rm g}$, meaning no hot corona at all, even though the data show the weak X-ray emission. We conclude that this result is obtained because, due to numerical reasons, the {\tt relagn} model is not able to reproduce such a weak hot corona. Therefore in the following we simply fix $R_{\rm hot}$ at $6\,R_{\rm g}$ and add the weak power law\footnote{In XSPEC, the power law is defined in energy space as a photon spectrum \( N_E \propto E^{-\Gamma} \), where $N_E$ is the photon flux density (photons/keV/$\text{cm}^2$/s), $E$ is the photon energy (keV), and $\Gamma$ is the photon index.} with fixed index of 1.8 to describe the X-ray data points. 
Letting the multi-colour disk component of the {\tt relagn} model to reproduce the optical-UV emission, the fit yields, with acceptable statistics, a slightly lower BH mass of $1.51 \times 10^{9}~ \text{M}_\sun$, but this solution stresses the disk energetics, leading to a higher accretion rate $\dot m$ of 0.174, and does not naturally produce the soft X-ray excess. This is our Model 1b.
The fitting parameters are shown in Table \ref{table:4} and the corresponding best-fit is displayed in the panel a) in Fig. \ref{fig9}. The plot includes the separate power law component modeling the X-ray data points.

\begin{figure}[h!]
   \centering
   \includegraphics[width=\hsize]{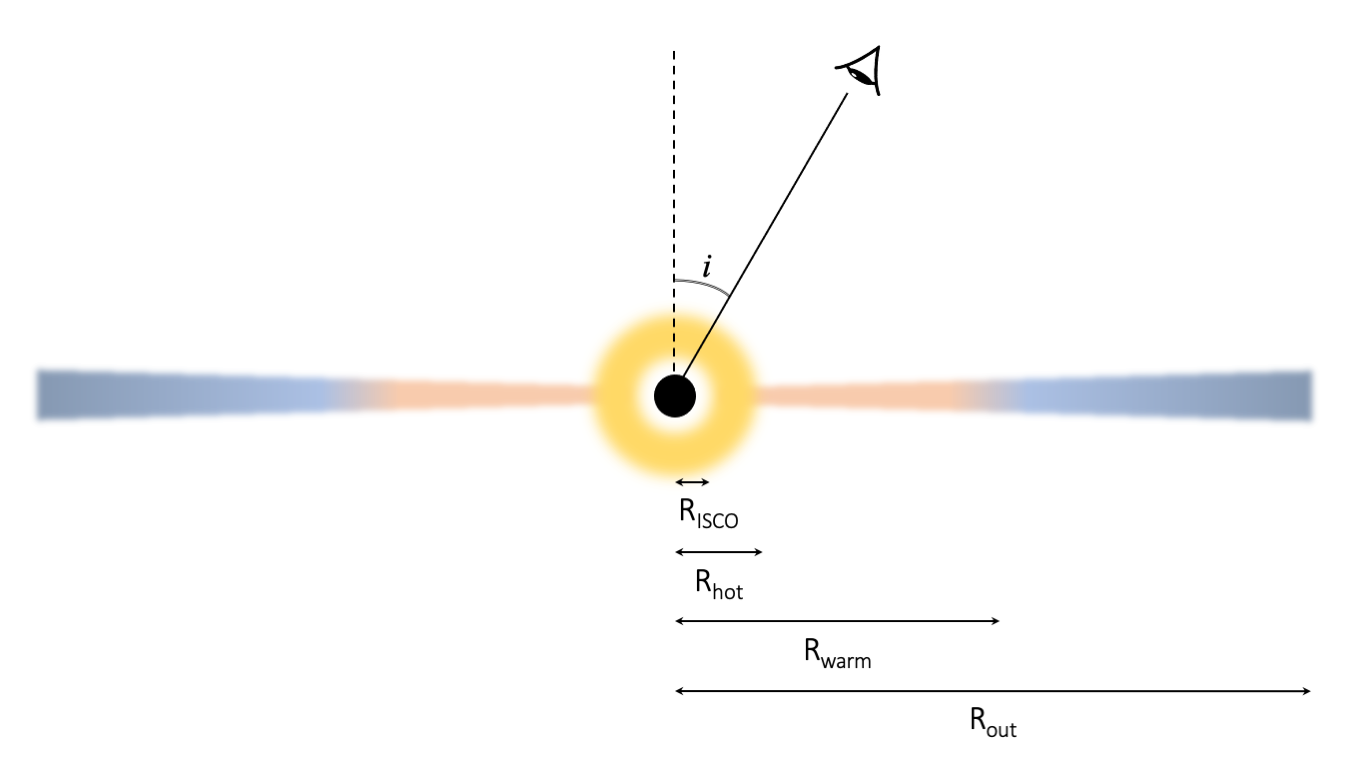}
      \caption{Schematic geometry used in the fitting process for Model 2. As in Fig. \ref{fig4}, the BH is in black, the standard accretion disk in blue-gray, and the compact and weak hot corona in yellow. The warm region is in light orange.}
         \label{fig10}
   \end{figure}

%
Secondly, we focused on the warm region. 
Although the results of the previous fit formally indicate that a warm region contribution is not required by the data ($\chi^2<1$), we decided to test for presence of a warm corona. The motivation comes from the fact that the spectral component attributed to a warm corona is quite often observed in spectra of high accretion rate AGN and X-ray binaries \citep{done2007, petrucci2018}.
This is our Model 2, illustrated in Fig. \ref{fig10}.
%

Allowing for warm corona in the {\tt relagn} model improves the fit considerably, with $\Delta\chi^2 = 8.69$ for three less degrees of freedom.
The best-fit warm region has \(kT_{\rm e,warm} = 0.19_{-0.09}^{+0.06}~ \text{keV}\), \( \Gamma_{\rm warm} = 3.83_{-0.19}^{+0.47} \), and \(R_{\rm warm} = 34_{-7}^{+10}~ R_\text{g}\), consistent with an optically thick warm corona extending to tens of gravitational radii and invoked to produce the soft X-ray -- UV excess. The best-fit results are reported in Table \ref{table:4} and plotted in Fig. \ref{fig9}, panel b).
So, Model 2 provides a significantly improved fit, although we do note that the improvement mainly comes from the two highest energy points in UV, around $1000\,$\r{A}, and there are no data points in far-UV -- soft X-ray range where this spectral component would be directly observable. 
To evaluate the influence of the 10-keV X-ray data point, we also performed a fit excluding it: the fit quality improved from $\chi^2$ = 3.93 for 12 dof to $\chi^2$ = 2.37 for 11 dof, as expected when removing the highest-energy constrain.

\section{The obscuration scenario and X-ray reflection}
\label{sec:obs}

The low level of X-ray emission in SDSS J101353 can {\it a priori} be explained in two ways. Firstly, it can be a very weak intrinsic emission from the central X-ray source, similar to spectral states observed in some soft X-ray transients. This spectral state is referred to as the ultra soft-state (USS) and occurs at an accretion rate not too high above the transition from hard state to soft state (see, e.g., \citealt{done2007} for a review).

\begin{figure}[h!]
   \centering
   \includegraphics[width=\hsize]{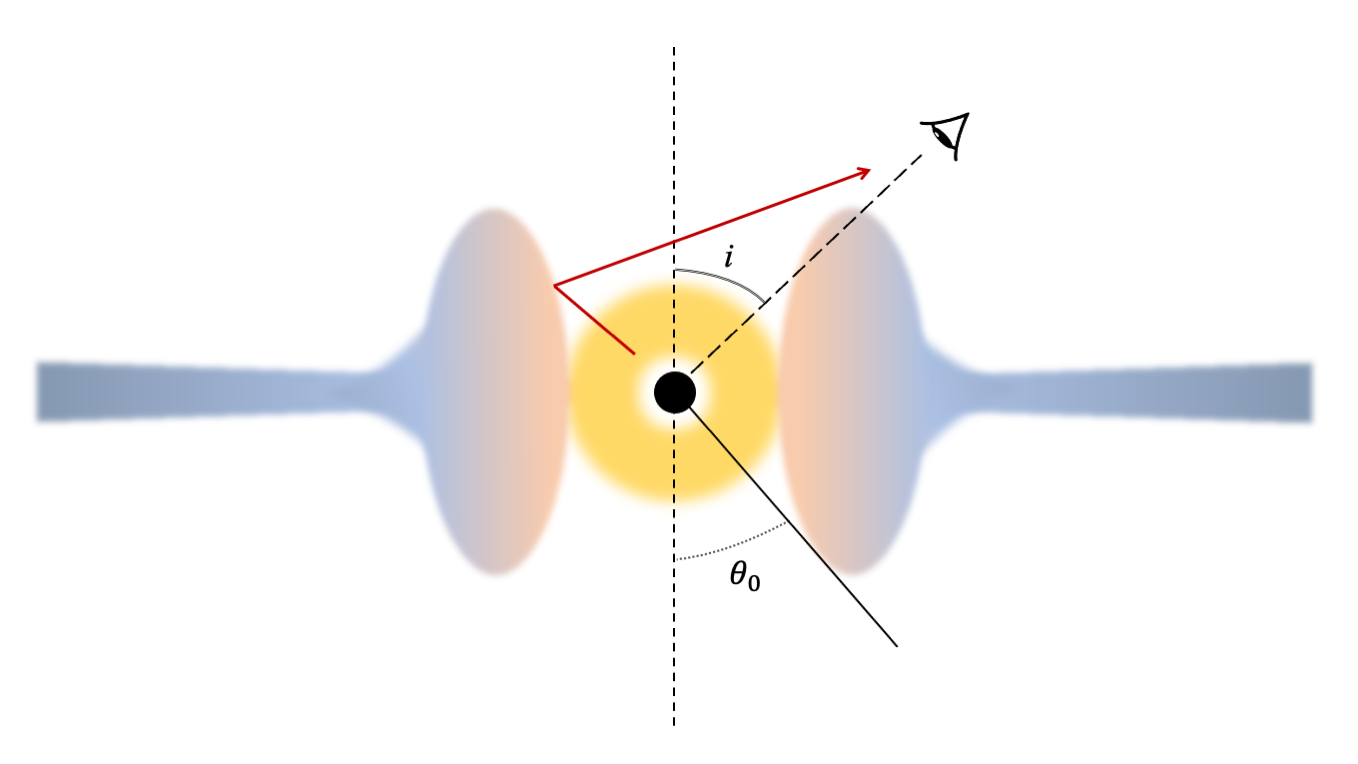}
      \caption{Schematic illustration of the obscuration scenario (Model 3) showing the emission reflected received by an observer with an inclination angle $i = 45^{\circ}$. The central BH and the hot corona are depicted as shown in Fig. \ref{fig10}. The gaseous obscurer is represented by the puffed up shape and has an opening angle $\theta_0$.}
         \label{fig11}
   \end{figure}

The second possibility is to assume that the central X-ray source is obscured and only a reflected/reprocessed component is reaching the distant observer, Model 3, see Fig. \ref{fig11}. This scenario, which is compatible with general ideas about broad range of X-ray emission in WLQ (recent paper \citealt{cheng2025}) assumes that the obscuration, that is invoked generally to reduce the high energy radiation reaching the BLR and explain the WLQ phenomenon, may also affect the emission directed toward the observer. Thus, the observed emission corresponds only to the reflected/reprocessed component while the intrinsic X-ray level is noticeably higher.

This scenario would require that the obscurer is, for example, geometrically extended to the lines of sight toward an observer and its optical depth is high enough to completely block the X-rays that, when redshifted to the observer's frame, would have been observed in the typical energy band of our X-ray detectors, $E>0.1$ keV.
We attempted to estimate the level of intrinsic emission in this scenario. We used a Monte Carlo code for radiative transfer in optically thick media that had previously been done in the context of obscuration in Seyfert 2 galaxies \citep[e.g.,][]{madejski2000}.

We consider a simple torus geometry with a rectangular cross-section, parameterized by its vertical and radial Thomson depth and by its half-opening angle. By assumption, the torus is Thomson very thick along the line of sight to the observer. Considering that quasars are expected to be viewed at low to moderate inclination angle, the opening angle cannot be too large.
In our baseline case, the torus has a half-opening angle of 20$^{\circ}$ and a square cross section with $\tau_{\rm T}$=10.

Monte Carlo simulations of photons isotropically emitted at the center of the system and transferred through this medium include the processes of photoelectric absorption and Compton scattering. The implemented atomic data allow for simulations of the fluorescent Fe K$\alpha$ line photons to be considered in the case of nonionized plasma.
We investigate two limiting scenarios: (i) a neutral torus with solar elemental abundances, and (ii) a pure-scattering case with no photoelectric absorption, as might approximate a hot, puffed-up inner disk obscuring structure (but note that we do not consider Compton up-scattering in this case).

\begin{figure}[h!]
   \centering
   \includegraphics[width=\hsize]{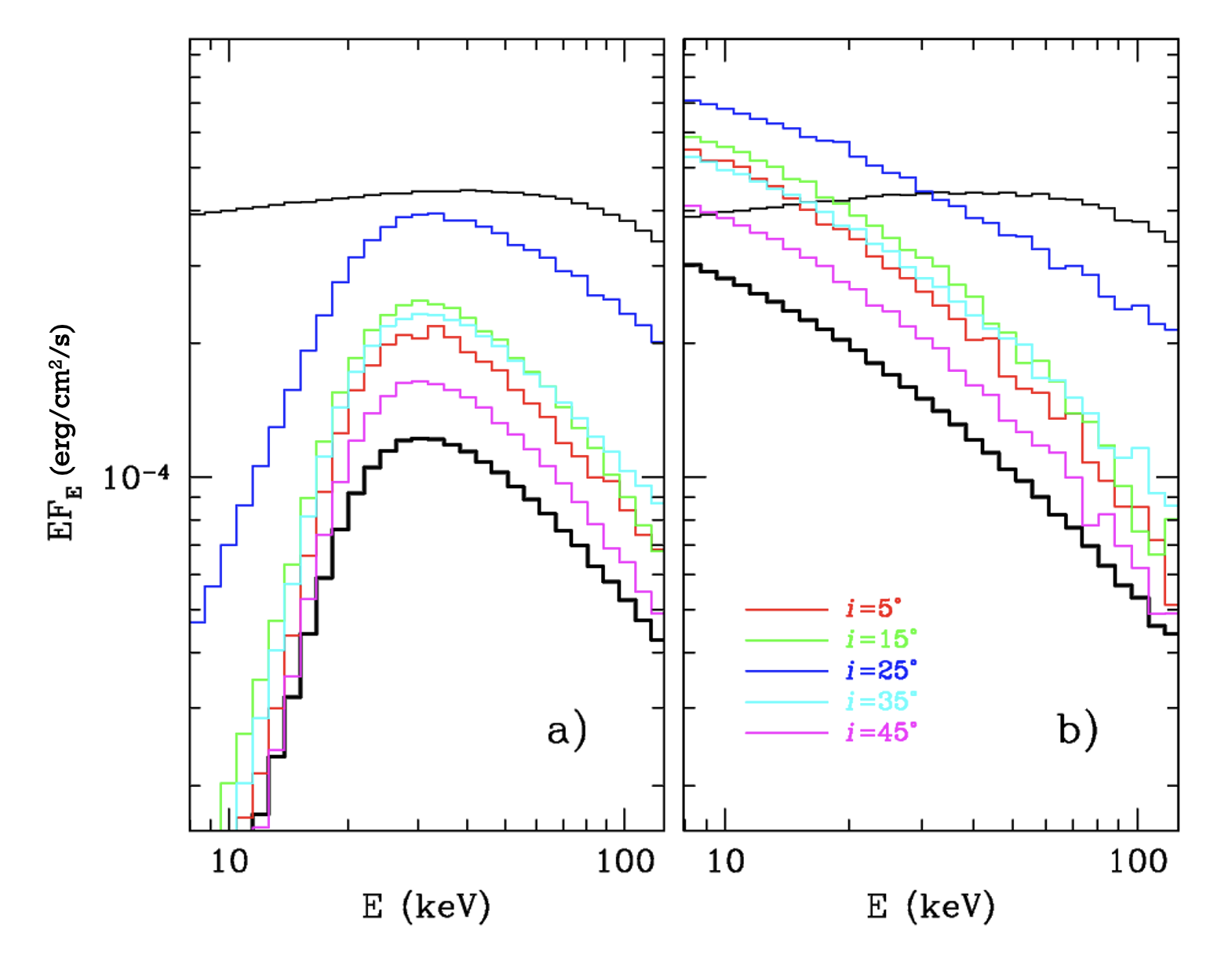}
      \caption{Spectra of the component reflected/reprocessed by the toroidal material obscuring the central source. Panel a) shows the case of cold, nonionized plasma, while panel b) is for the fully ionized case (but note that Compton up-scattering is not taken into account). The black thin line is the primary emission (assumed a thermal Comptonization for \(kT_e = 100~ \text{keV}\)), the black thick line shows the angle averaged reflected component, while the colored lines show the dependence of the reflected spectra on the viewing angle, $i$, for $i<50^{\circ}$.}
         \label{fig12}
   \end{figure}

\begin{figure}[h!]
   \centering
   \includegraphics[width=\hsize]{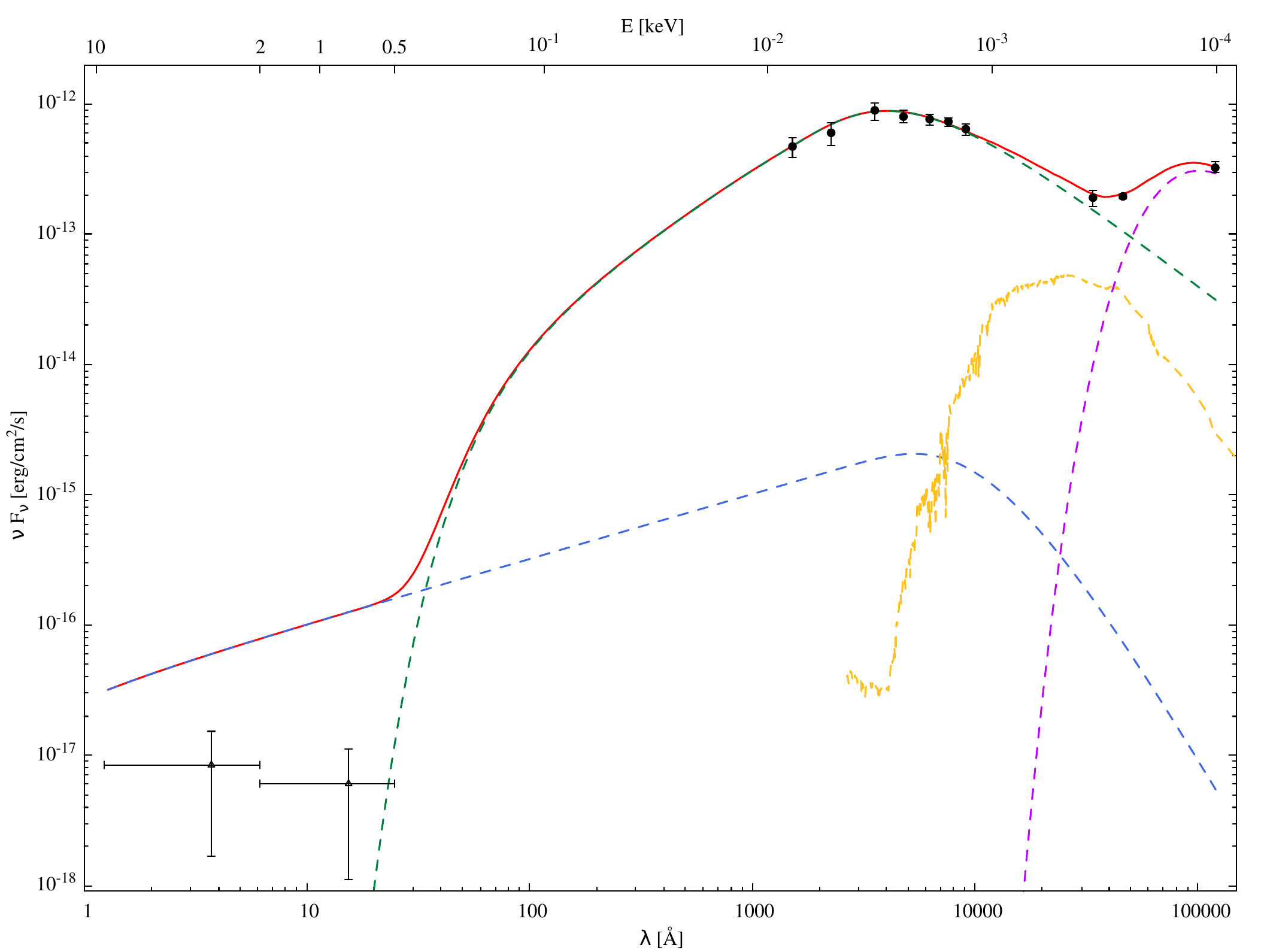}
      \caption{Broadband SED fit of SDSS J101353 with the estimated primary emission (dashed blue curve) using \texttt{pexrav} and \texttt{nthcomp} in Model 3. As previously, the total model, the \texttt{relagn} component, the starlight and torus contributions are shown in red, green, yellow and violet, respectively. The data are displayed with the black dots.}
         \label{fig13}
   \end{figure}

The results presented in Fig.~\ref{fig12} show unfolded spectra where $F_E$ is the energy flux density per unit energy. The spectrum is shown as $EF_E$ to illustrate the energy at which the broadband SED peak occurs. The spectra indicate that the reflected hard X-ray spectrum is not significantly suppressed compared to the intrinsic emission. 
At $\sim$ 30 keV, the ratio of primary to reflected flux is not higher than 4 for viewing angles of $\sim$ 40$^{\circ}$ (see the magenta histogram).
Having estimated the primary emission in this way, we can now construct the spectrum reaching the BLR, i.e., the intrinsic spectrum. For this, we model the data replacing the power law with the \texttt{pexrav} model (representing the reflected component, see \citealt{magdziarz1995}), and using the \texttt{nthcomp} model to represent a thermally Comptonized continuum (\citealt{zdziarski1996, zycki1999}), to fit to the hard X-ray points, and then adjust the amplitude of the required primary emission so that the ratio of fluxes at 10 keV (approximately 30 keV redshifted to the observer frame) is equal to 4. \texttt{nthcomp} thus provides the primary continuum emission. This is represented in Fig.~\ref{fig13}, where the resulting spectrum is plotted.
The results of our spectrum inspection and our SED fitting highlight both the extreme X-ray weakness and the lack of prominent emission lines in SDSS J101353, raising important questions about the physical processes at play. In the following discussion, we consider several possible scenarios that could explain these peculiar properties.

\section{Discussion}
\label{sec:discus}

The WLQ SDSS J101353 combines two properties that place it among the most extreme WLQs: a nearly featureless UV-optical spectrum, with only a weak Mg II emission line (while both C III] and C IV are almost absent), and a pronounced X-ray weakness, $\sim 4$ orders of magnitude below expectations for normal quasars. 
This dual deficit challenges standard quasar models and requires considering both accretion physics and BLR conditions. 
The quasi-absence of high-ionization emission lines may result from a combination of factors: a sparsely populated BLR, shielding of ionizing photons, or a fundamentally different X-ray SED \citep{shemmer2009, luo2015}. 
Nevertheless, in this article, we do not consider the production of emission lines by the BLR or the presence of the BLR itself. Our analysis focuses exclusively on the continuum emission. More precisely, our fitting constrains the thermal emission from the accretion disk, the presence of a parsec-scale torus, and the starlight from the host galaxy. These components alone, however, cannot account for the suppression of the emission lines.

We confirm that the shape of the IR/optical continuum of SDSS J101353 is standard, 
lying within the range of shapes observed among other type 1 quasars (the photon index of our WLQ $\alpha_{\nu}= -0.46$). However, the low level of X-ray radiation is puzzling. It can be explained in both ways. One option is that the central X-ray emitter is intrinsically faint. Alternatively, the weak signal may result from heavy obscuration of the primary X-ray source.

Investigating the first scenario, we model the continuum of SDSS J101353 using both \texttt{kerrbb} and \texttt{relagn} models (in XSPEC), which provides an independent and physically motivated estimate of the accretion flow. 
The best-fits of both models suggest that the thermal emission extends down to the innermost stable circular orbit (ISCO), consistent with a standard thin disk. A disk-dominated  configuration with an exceptionally weak hot corona (which is inefficient, very compact, or partially quenched) is therefore favored. The hot corona can only generate a very small amount of energy, to account for the observed X-ray data. 
Nevertheless, the inferred coronal parameters themselves are comparable to those reported for typical quasars. Its photon index $\Gamma = 1.8$.

The weak emission lines observed in SDSS J101353 are difficult to interpret. 
\citet{LD2011} proposed that an inefficient cold accretion disk caused by a very massive BH could generate a nonionizing continuum. In their model, the accretion disk becomes too cold to ionize when $M_{\rm BH} > 3.6 \times 10^9\ \text{M}_\sun$ for a nonrotating BH. 
This threshold is derived under the assumptions that the typical X-ray luminosity observed in quasars satisfies $L_X \lesssim L_{\rm bol}$ and that the ionizing luminosity of a cold accretion disk is only $L_{\rm ion} \sim 0.01 L_{\rm bol}$ (where $L_{\rm bol}$ is the bolometric luminosity). For SDSS~J101353, in case of Models 1a/1b, we got $\log (L_X/L_{\rm bol}) = -4.87$ and  $L_{\rm ion}/L_{\rm bol} \simeq 0.002$, and the supermassive BH may be lighter than the aforementioned upper limit.
From the single-epoch Mg II line, we found a BH mass of \( \sim 5 \times 10^8\ \text{M}_\sun \), typical of normal quasars. Nevertheless, these estimates are subject to biases affecting WLQs. \cite{marculewicz2020} showed that virial masses derived from the FWHM of H$\beta$ can be underestimated by factors of 4-5 in WLQs, compared with disk-fitting techniques, due to geometric or dynamical differences in their BLRs. If a similar bias affects the Mg II line, then the true mass of SDSS J101353 may be significantly higher than the single-epoch estimate.
Indeed, our continuum modeling using the disk-fitting methods (both \texttt{kerrbb} and \texttt{relagn} models) provides a physically motivated, BLR-geometry-independent estimate of the accretion flow. Both methods converge toward a larger BH mass, of \( \sim (1.5-2.1) \times 10^9\ \text{M}_\sun \), but still smaller than the upper limit found by \citet{LD2011}.

The inclusion, in our \texttt{relagn} fit (i.e., Model 2), of a warm, optically thick Comptonizing region substantially improves the fit.
From this warm-corona fit (\(kT_e \approx 0.20~ \text{keV},~ \Gamma \approx 3.80\)) and using the definition of $\Gamma$ in \cite{beloborodov1999}, we found that the warm corona has an optical depth of \(\tau \approx 7.26\). 
The presence of a warm Comptonizing region supports a configuration seen in many AGN (e.g., \citealt{kubota2018, petrucci2018, middei2020}), although the limited X-ray data prevent strong constraints. These works show that such a warm Comptonizing layers are a robust explanation for the soft X-ray excess and line production in quasars. Our results suggest that SDSS J101353 shares similar physical conditions with them, despite its extreme weakness in the hard X-ray band. 
In Model 2, we obtained $\log (L_X/L_{\rm bol}) = -4.29$ and  $L_{\rm ion}/L_{\rm bol} \simeq 0.05$, both of which are still low.

This suggests that the X-ray level and the reduced coronal contribution are intrinsic, indicating that the corona is either physically absent or unusually compact. 
The quasar continuum resembles the spectral behavior observed in certain soft X-ray transients -- a subclass of X-ray binaries. These transients typically appear when the accretion rate is only slightly above the threshold that separates the hard and soft states.
As shown by \citet{done2007}, the transitions from the hard to the ultrasoft state (USS) are accompanied by dramatic changes in the geometry of the inner accretion flow, particularly a strong suppression or disappearance of the hot Comptonizing corona (see their figure 9). In the ultrasoft state, the spectrum becomes almost entirely disk-dominated, with only a faint high-energy tail.  
This configuration closely resembles the SED of SDSS J101353. If this analogy holds, then this WLQ may represent an AGN version of a low-corona state, naturally suppressing the production of high-energy photons and the ionizing continuum required to power high-ionization emission lines. This provides a unified explanation for the simultaneous weakness of emission lines and the X-ray deficit.

Regarding the X-ray weakness, we propose a second explanation of this phenomenon
seen in SDSS J101353. While the ultrasoft state analogy offers a compelling physical picture, geometric effects may also contribute. The production of a shielding gas or a puffed-up inner disk between the inner region and the observer (Model 3) means that the observer only detects radiation that has been reflected or reprocessed.
The X-ray level is lower than typical and the shielding gas attenuates the ionizing continuum reached by HILs while still allowing the outer BLR to receive some radiation 
by LILs \citep[e.g.,][]{wu2011, gallo2011, cheng2025}. Our SED fits cannot rule out such a configuration. Therefore, we tested a reflection-dominated X-ray flux that could mimic the observed faintness. The resulting estimated primary X-ray flux is still $10^2$ - $10^3$ times fainter than that of a typical quasar.
The explanation for the emergence of the puffed-up inner disk is the slim disk solution
\citep{ACLS88,Czerny2019}. It requires a high accretion rate (sub- or super-Eddington). 
From the single-epoch Mg II line, we found an Eddington accretion rate of 0.17, typical of normal quasars. The corresponding accretion rate is modest compared to the range values of 0.3-0.6 reported for WLQ samples \citep{marculewicz2020}. In contrast, \citet{cheng2025} argue that WLQ accretion rates may be underestimated in single-epoch methods, especially when the EW of C IV decreases. Notably, their figure 12 shows that the Eddington ratio distribution of WLQs (at redshifts from 1.45 to 1.90) is not dramatically higher than for normal quasars. Their calculated mean Eddington ratios are $\dot m =0.151$ for WLQs, and $\dot m =0.119$ for normal quasars. This indicates that the high-Eddington WLQ picture may not apply uniformly across all redshifts and all WLQ subclasses. Our result for SDSS J101353 fits naturally into this more nuanced view. Additionally, our continuum modeling using XSPEC models for the disk fitting provides a still moderate Eddington accretion rate \( \dot{m} \sim 0.1 \) in this WLQ. 
This may suggest a greater likelihood of SDSS J101353 being in the USS state than of the presence of the shielding gas or puffed-up inner accretion disk.

According to the Mg II emission line fit, we know that the FWHM of this line is $3875 \pm 250$ km s$^{-1}$. In addition, its UV iron-to-magnesium ratio $R_{\rm Fe II, UV}$, defined as \( \mathrm{EW(Fe II_{UV}(\lambda2700-2900\AA))/ EW(Mg II)} \), lies in the range 5.5-7.9. These values were derived using the relations between $\mathrm{FWHM(Mg II), EW(Mg II)}$ and $\mathrm{EW(FeII_{UV})}$ for quasars \citep{sniegowska2020}.
It is worth mentioning that these values classify our WLQ as a member of the A population of the quasar main sequence. Sources belonging to this population are characterized by higher accretion rates (e.g. $\log\dot m \gtrsim -1.3$ in Marziani's sample), narrower lines, a higher intensity Fe II to Mg II ratio, blueshifted C IV, and the presence of a strong wind \citep{marziani2024}.

All above conclusions are further confirmed by the value of so called X-ray loudness i.e., $\aox$ parameter \citep{2009sobolewska}, that represents the ratio of optical to X-ray luminosity weighted by the wavelength for which the both luminosities have been calculated. In case of the data collected of our special quasar, and for 
two energy points 2500 \AA\ and 2\,keV\footnote{\( \aox = -0.3838 ~\text{log} ~[L(2 ~\text{keV}) / L(2500 ~\AA )]\). We use the minus sign, therefore our $\aox$ parameter is positive.}, we obtained $\aox=2.06$. Together with the derived Eddington accretion rate, it puts SDSS~J101353 in the top, right corner of the $\aox$ vs $\dot m $ plane, presented for quasars by \citet{ruan2019} (see their figure 3). This finding fully confirms conclusions of this paper that the source is in its high/soft spectral state, where the corona is confined to the compact inner hot region.

The weak corona might also be episodic. Indeed, another possibility is that SDSS J101353 is undergoing a changing-look, or state-change phase. A reduced coupling between the disk and corona could lead to a transient phase characterized by diminished X-ray emission and suppressed high-ionization lines. Such a behavior has been reported in nearby AGN, where quasars have transitioned from X-ray bright to X-ray weak states, sometimes accompanied by spectral changes in the BLR (e.g., \citealt{noda2018, ruan2019}). If SDSS J101353 is caught during such a weak-corona state, this could simultaneously explain its current X-ray faintness and line suppression. Recent observations of a 're-awakening' in the AGN Markarian 590 \citep{palit2025} show that a rising continuum, including a re-emerging warm corona and increased soft X-ray/UV flux, is accompanied by the return of broad Balmer and high-ionization lines.

\section{Conclusions}
\label{sec:concl}

We have performed a detailed broadband spectral analysis of the WLQ SDSS J101353, a quasar that combines two remarkable features: a nearly featureless optical-UV continuum, with only a faint Mg II line detected, and an exceptionally weak X-ray emission.
We have focused on its continuum and X-ray properties. Our main conclusions are summarized as follows:
\begin{itemize}
    \setlength\itemsep{1em}
    \item Robust continuum fitting: both the relativistic thin-disk model \texttt{kerrbb} combined with a power law, and the multi-component \texttt{relagn} model yields consistent black hole parameters ($\text{M}_\text{BH} \sim 2 \times 10^9 ~\text{M}_\sun$, $\dot m \sim 0.1$), confirming that SDSS J101353 accretes at a typical quasar rate. However, the coronal emission is markedly suppressed.
    \item Evidence for a warm corona: including a warm, optically thick Comptonizing region (\(kT_{e,warm} \approx 0.20~ \text{keV}\), \(\Gamma_{warm} \approx 3.8\), and $\text{R}_{warm}$ extended to 34 $\text{R}_\text{g}$) significantly improves the fit, pointing toward an inefficient or compact corona as the primary cause of the X-ray faintness.
    \item Additional mechanisms: we investigate additional physical mechanisms, such as a shielding by a thick inner disk, or a partial absorption of ionizing photons. If we assume that we only see the X-ray emission reflected by the shielding gas, the primary X-ray level remains low, no more than 4 times the observed one. It indicates the intrinsic weakness of the hot corona.
    \item Implications for WLQ diversity: SDSS J101353 reinforces the view that WLQs constitute a heterogeneous population in which both structural and temporal effects can suppress the production of high-energy photons and broad emission lines. 
    This source may represent an AGN version of the ultrasoft state (USS) seen in some 
    soft X-ray transients.
    This spectral state occurs in XRBs at an accretion rate not too high above the transition from hard state to soft state.
\end{itemize}

Taken together, our results suggest that SDSS J101353 offers a valuable counterexample to the idea that all WLQs are extreme high-Eddington accretors.

Future work (paper II) will explore detailed photoionization simulations using Cloudy, testing whether the observed emission line weakness can be reproduced under different continuum and geometry assumptions. 
Multi-epoch observations could also be essential to determine whether the X-ray weakness of SDSS J101353 reflects a stable configuration or a transient state of the accretion flow.

\begin{acknowledgements}
The authors thank Marianne Vestergaard for providing the iron emission template, and Profs. Czerny and Done for helpful discussions and advice. They also thank the anonymous referee for constructive comments that improved the manuscript.
The research leading to these results has received funding from the European Union's Horizon 2020 Programme under the AHEAD2020 project (grant agreement n. 871158). RW has been fully and AR has been partially supported by the Polish National Science Center grant No. 2021/41/B/ST9/04110.
\end{acknowledgements}

\bibliographystyle{aa} 
\bibliography{biblio.bib} 

\end{document}